\documentstyle [12pt,epsf]{article}

\input{epsf}
\begin{document}
\begin{titlepage}
\begin{center}
  
 \vspace{-0.7in}

{\large \bf The Partition Function for the Anharmonic Oscillator\\ 
in the Strong-Coupling Regime}\\
\vspace{.5in}{\large\em N. F. Svaiter\,\,\footnotemark[1]}\\ 

Centro Brasileiro de Pesquisas Fisicas-CBPF\\ Rua Dr. Xavier Sigaud 150,
Rio de Janeiro, RJ, 22290-180, Brazil\\

\subsection*{\\Abstract}
\end{center}

We consider a single 
anharmonic oscillator with frequency $\omega$ and 
coupling constant  $\lambda$ respectively, in the strong-coupling regime.
We are assuming that the system is in thermal equilibrium with a 
reservoir at temperature $\beta^{-1}$. Using the strong-coupling perturbative
expansion, we obtain the partition function 
for the oscillator in the regime $\lambda>>\omega$, 
up to the order $\frac{1}{\sqrt{\lambda}}$.
To obtain this result, we use of a combination of Klauder's
independent-value generating functional
(Acta Phys. Austr. {\bf 41}, 237 (1975)),
and the generalized zeta-function method. The free energy and 
the mean energy, up to the order $\frac{1}{\sqrt{\lambda}}$, are also presented. We are showing that the thermodynamics
quantities are nonanalytic in the coupling constant.

\footnotetext[1]{e-mail:nfuxsvai@cbpf.br}

PACS numbers:03.70+k,04.62.+v

\end{titlepage}
\newpage\baselineskip .37in
\section{Introduction}
The strong-coupling regime in quantum field theory is one of the unsolved
problems of theoretical physics of the last century. There are many 
situations where one has to account for non-perturbative coupling 
regions, and to discuss the physics of the 
strongly coupled systems. Let us quickly 
mention a few examples.

The standard case is QCD, where 
we believe there exists a deconfinement phase transition. It is 
expected that at sufficiently high temperatures, quarks and gluons 
become no longer bounded inside hadrons. There is also another 
phase transition which depends on the number of quark flavour: 
the chiral symmetry phase transition. The chiral symmetry is spontaneoulsly
broken at zero temperature and it is restored at some finite 
temperature. Although QCD is asymptotically free in the 
ultraviolet limit, it has a coupling $g(\beta^{-1})\approx 1$ near the 
phase transition, where $\beta^{-1}$ is the temperature.
Consequently a non-perturative approach 
must be used to deal with this situation. 
The most fruitful idea is lattice QCD which is a means 
to study the phase transition from first principles.

In condensed matter physics we also have different situations
in which the strong-coupling regime may occur. For example,
in the 
theory of superconductors there is the strong-coupling 
regime in the electron-phonon interaction. In particular, the 
strong-coupling limit for the polaron problem has been 
investigated by many authors. A detailed exposition can be found in Ref. \cite{polaron}. Also in models with a 
disordered ground state one can use the strong-coupling approach. 
A standard case is the $O(N)$ non-linear sigma 
model, describing lattice magnets, where there is a regime in which
one can treat the coupling between 
different sites as a perturbation.
Until now we pointed out  
that there are many situations in quantum field theory that 
cannot be described using the weak-coupling perturbation theory. 
Nevertheless, the use of the perturbative expansion 
outside the weak-coupling regime is also fundamental 
for our understanding of the 
whole perturbative renormalization program in quantum field theory.

Various types of harmonic oscillator have served for long as 
simple analogue systems for more general and complex situations
in quantum field theory. Actually, 
the anharmonic oscillator, with a $\frac{\lambda}{4!}
\,x^{4}(\tau)$ term is formally very similar to the field theory 
describing a scalar field with a quartic self-interaction.
Therefore, in this paper we present 
a method for calculating the partition function and the Helmholtz 
free energy for a single oscillator with 
the anharmonic $\frac{\lambda}{4!}
\,x^{4}(\tau)$ contribution in the strong-coupling regime 
up to the order $\frac{1}{\sqrt{\lambda}}$. In 
other words, our model belongs to the regime in which 
the coupling constant is much larger than the frequency, i.e., 
$\lambda\,>>\omega$. This is also a convenient way to study the regime of
large fluctuations. In fact, the oscillator with 
any anharmonic term of the kind
$\frac{\lambda}{(2p)!}\,x^{2p}(\tau)$,\,\,$p>3$ in the
strong-coupling regime can be analysed by our method and also 
the generalization for $N$ non-interacting anharmonic oscillators is
straightforward.

The anharmonic oscillator is
described by the Hamiltonian
\begin{equation}
H=\frac{p^{2}}{2m}+\frac{1}{2}\,
m\,\omega^{2}\,x^{2}+\frac{\lambda}{4!}\,x^{4},
\label{1}
\end{equation}
and for simplicity we are assuming that our system is 
one-dimensional and is in thermal equilibrium with a reservoir 
at temperature $\beta^{-1}$. We are working in the imaginary time 
formalism and making use of the Kubo-Martin-Schwinger (KMS) condition
\cite{kms1} \cite{kms2} \cite{kms3}.

To find the partition function and the Helmholtz 
free energy, our approach consists in the 
combination of two techniques used currently in the 
literature: the strong-coupling expansion \cite{kovesi} \cite{be1} 
\cite{parga} \cite{be2} \cite{co} \cite{be3}
and the zeta-function method  \cite{seeley}
\cite{hawking}, which is a classical case of a general formalism concerning 
regularized products and determinants. For a
review of the strong-coupling expansion in field theory, 
see for example Ref. \cite{nami1} and also Ref. \cite{nami2}, and 
for a recent treatment of the strong-coupling expansion in quantum 
mechanics, see for example Ref. \cite{A}.

Let us briefly discuss the strong-coupling expansion in Euclidean 
field theory at zero temperature.  
The basic idea of the approach is the following: 
in a formal representation for the generating functional of 
complete Schwinger functions of the theory $Z(h)$, we treat
the Gaussian part of the action as a perturbation with 
respect to the remaining terms of the functional integral, 
i.e., in the case for the  $(\lambda\varphi^{4})_{d}$ theory, the 
local self-interacting part, in the 
functional integral. In the literature 
this approach it has been called the strong-coupling expansion. 
A stimulating reading can be found in Ref. \cite{jrk} and Ref. \cite{rivers}.
The main difference from the the standard perturbative expansion is that 
we have an expansion of the generating functional of complete Schwinger functions
in inverse powers of the coupling constant. We are developing our perturbative
expansion around the independent-value generating functional
$Q_{0}(h)$, where
different points of the Euclidean space are decoupled, 
since the gradient terms are dropped \cite{ca} 
\cite{sol1} \cite{kla} \cite{meni}.  For study the 
analytic structure of the independent-value 
generating-functional in the complex plane of the coupling constant $\lambda$, see for example Refs. \cite{baker} and \cite{eu}.

In the perturbative expansion around the independent-value model,
the representation of the Schwinger functional $Z(h)$ is defined 
by an unrenormalized perturbative 
series, which  can be truncated in the order of the approximation.
For example, if $Z(h)\equiv Q_{0}(h)$, we call it the independent-value
approximation or the zeroth-order approximation. 
It should be noted that 
although the independent-value functional $Q_{0}(h)$ is not 
a product of Gaussian integrals, it can be viewed formally as an 
infinite product of ordinary integrals, one for each point of the 
d-dimensional Euclidean space. 
The fundamental problem of the strong-coupling 
expansion is how to give meaning 
to the independent-value generating functional and to the unorthodox 
representation for the Schwinger functional.

A solution to this 
problem, was presented by Klauder and colaborators a long time ago
\cite{sol1} \cite{sol2} \cite{kla2}. They obtained a quite interesting
expression for the independent-value generating functional describing 
Euclidean free and also self-interacting scalar fields. We would like 
to stress that a naive use of a continuum limit of the lattice 
regularization for the independent-value generating functional leads to a 
Gaussian theory, where we simple make use of the central limit theorem.
The fundamental modification which allow us to avoid the central limit 
theorem is a change in the measure in the functional integral. 
Consequently, in the Klauder's derivation, starting from the 
independent-value generating functional, and taking
the limit where the coupling constant becomes null, the interacting 
theory solutions become the pseudo-free one.

In this paper we show how it is possible to compute the 
partition function and the Helmholtz free energy of the anharmonic oscillator
in the strong-coupling regime, 
up to the order $\frac{1}{\sqrt{\lambda}}$. 
Our results are showing that the thermodynamics
quantities are nonanalytic in the coupling constant.
We would like to stress that although, by a unitary transformation 
the quartic anharmonic oscillator can be expressed in a form where 
the dependence on the coupling constant appears in an overall 
factor, we are using the standard form, since we are interested
in the regime where the anharmonic terms are much larger than the 
kinetic term. In other words we are interested in the regime 
of large fluctuations.

This paper is organized as follows: In section II, 
the strong-coupling expansion for 
a single anharmonic oscillator is presented.
In section III we calculate the 
partition function and other thermodynamics
quantities that are derived from it, as the 
Helmholtz free energy and the mean energy of the system.  
Finally, section IV contains our conclusions. 
We assume that the physical quantities are dimensionless.
Consequently, it is convenient to introduce 
an arbitrary parameter $\mu$ with mass 
dimension to define all dimensionless physical quantities. 
For simplicity, we assume that $\mu=1$ since we are not interested
in the scaling behavior of the model.   
As usual $\beta^{-1}$ is the temperature of the thermal bath.

\section{The partition function of a one-dimensional quantum mechanical 
system}\

Let us consider a one-dimensional quantum mechanical system. The 
partition function for the system assuming that it is in thermal 
equilibrium with a reservoir at temperature $\beta^{-1}$ is given by
\begin{equation}
Z(\beta)=\int_{x(0)=x(\beta)}[dx(\tau)]\,\,\exp\Biggl[-\int_{0}^{\beta}
d\tau\left(\frac{1}{2}m(\frac{dx}{d\tau})^{2}+V(x(\tau))\right)\Biggl],
\label{2}  
\end{equation}
where in the functional integral we require that $x(\tau)$ is periodic
with period $\beta$, i.e., $x(\tau)=x(\tau+\beta)$. 
There are many different physical situations that 
can be analysed starting from the partition function. We would like 
to discuss the case of a single anharmonic oscillator, where 
the contribution of $V(x(\tau))$ is given by
\begin{equation}
V_{1}(x)=\frac{1}{2}\,\omega^{2}\,x^{2}+\frac{\lambda}{4!}\,x^{4}.
\label{3}
\end{equation}
For simplicity, we are choosing $m^{2}=1$. 
As we will see, it is not difficult to apply our method in non-polynomials theories. The second situation 
that can also be analysed 
is the non-polynomial model, defined by the following expression
\begin{equation}
V_{2}(x)=\frac{\omega^{4}}{\lambda}\Biggl[
\cosh \left(\frac{\sqrt{\lambda}}{\omega}\, x(\tau)\right)-1\Biggl].
\label{4}
\end{equation}

In order to study the anharmonic oscillator, let us sketch the solution 
for the single harmonic oscillator. In the potential $V_{1}(x)$, choosing 
$\lambda=0$ we obtain the harmonic oscillator, where the partition 
function can be found in texbooks. The partition function can be written
in a more tractable way, where an integration by parts is done and 
a surface term is disregarded. To avoid minor complications, we 
will assume the following expression to the free partition function 
$Z(\beta)|_{\lambda=0}$. In particular, we shall therefore start from 
\begin{equation}
Z(\beta)|_{\lambda=0}=\int_{x(0)=x(\beta)}[dx(\tau)]\,\,\exp\Biggl[-
\int_{0}^{\beta}d\tau\,\frac{1}{2}x(\tau)\left(-\frac{d^{2}}{d\tau^{2}}+
\omega^{2}\right)
x(\tau)\Biggl].
\label{6}  
\end{equation}
It is a standard procedure to define the following 
kernel $K(\omega;\tau-\tau')$ by the equation
\begin{equation}
K(\omega; \tau-\tau')=\left(-\frac{d^{2}}{d\tau^{2}}+\omega^{2}\right)
\delta(\tau-\tau'),
\label{7}  
\end{equation}
and substituting Eq.(\ref{7}) in Eq.(\ref{6}) the free partition function 
$Z(\beta)$ becomes 
\begin{equation}
Z(\beta)|_{\lambda=0}
=\int_{x(0)=x(\beta)}[dx(\tau)]\,\,\exp\Biggl[-
\int_{0}^{\beta}d\tau \int_{0}^{\beta}d\tau'
\,\frac{1}{2}x(\tau)K(\omega;\tau-\tau')x(\tau')\Biggl].
\label{8}  
\end{equation}
As usual, we define the generating functional 
$Z(\beta; h)$ introducing an external source $h(\tau)$. At this point it is convenient to 
consider $h(\tau)$ to be complex. Consequently
$h(\tau)=\mbox{Re}(h)+i\,\mbox{Im}(h)$. In the paper we are 
concerned with the case $\mbox{Re}(h)=0$.
Therefore the generating functional $Z(\beta;h)$ is defined by 
\begin{equation}
Z(\beta;h)|_{\lambda=0}=\int_{x(0)=x(\beta)}
[dx(\tau)]\exp\Biggl[-
\int_{0}^{\beta}d\tau\int_{0}^{\beta}d\tau'
\frac{1}{2}x(\tau)K(\omega;\,\tau-\tau')x(\tau')+
\int_{0}^{\beta}d\tau\,h(\tau)x(\tau)\Biggl].
\label{89} 
\end{equation}
Note that we are using the same notation for functionals and functions, 
for example $Z(\beta;h)$ instead of the usual notation  $Z_{\beta}[h]$.
Since the integrations that appear in Eq.(\ref{89}) are Gaussian 
it is straightforward to write 
\begin{equation}
Z(\beta;h)|_{\lambda=0}=Z_{\beta}\,\exp\Biggl[-
\int_{0}^{\beta}d\tau \int_{0}^{\beta}d\tau'\,
\frac{1}{2}h(\tau)G(\omega;\,\tau-\tau')h(\tau')\Biggl],
\label{891} 
\end{equation}
where the partition function is defined by 
$Z(\beta)=Z(\beta;h])_{h=0}$, and the Green function
$G(\omega;\,\tau-\tau')$ is the 
inverse kernel, defined by
\begin{equation}
\int_{0}^{\beta}\,d\tau'\,K(\,\omega,\tau-\tau')G(\,\omega,\tau'-\tau'')=
\delta(\tau-\tau'').
\label{892} 
\end{equation}
At this point it is important to define the modified kernel 
$K(\omega,\sigma;\tau-\tau')$ by the equation
\begin{equation}
K(\omega,\sigma;\tau-\tau')=\left(-\frac{d^{2}}{d\tau^{2}}+
(1-\sigma)\,\omega^{2}\right)
\delta(\tau-\tau'),
\label{9}  
\end{equation}
where $\sigma$ is a complex parameter 
defined in the 
region $0\leq\,\mbox{Re}\,\,(\sigma)< 1$. Note that in general, 
$\mbox{Re}\,\,(\sigma) \neq 1$, 
because $\mbox{Re}\,(\sigma)=1$ introduce infrared divergences
in the calculations.
The zero frequency case $(\sigma=1)$ in the modified 
kernel can be assumed in some very special 
situations only, as for example to calculate the renormalized vacuum energy
of a scalar field in the presence of boundaries, where Dirichlet boundary conditions are assumed. See for example Ref. \cite{nami2}. 
To summarize, the choice of a suitable  $\sigma$ will simplify our 
calculations in some situations.
This modification will be clarified in the next section.

Let us now suppose an anharmonic oscillator in the strong-coupling regime.
We would like to stress that the semiclassical approximation
(the WKB approximation) can also be used in the strong-coupling 
regime. The difficulty is to find the classical orbits which are 
stationary-phase points in the functional integral \cite{C}. 
First, as we made in the non-interacting case, 
it is convenient to couple linearly the anharmonic
oscillator to a $\tau$-dependent external 
source. Therefore the generating functional at finite 
temperature is given by
\begin{eqnarray}
&& Z(\beta;h)=\int_{x(0)=x(\beta)}[dx(\tau)]
\nonumber\\
&&\exp\Biggl[-
\int_{0}^{\beta}d\tau\int_{0}^{\beta}d\tau'\,
\frac{1}{2}x(\tau)\,K\,x(\tau')+
\int_{0}^{\beta}d\tau\left(-\frac{1}{2}\sigma\omega^{2}x^{2}(\tau)
-\frac{\lambda}{4!}x^{4}(\tau)+h(\tau)x(\tau)\right)\Biggl],
\label{10} 
\end{eqnarray}
where we have also integrated over all periodic paths and  
$K\equiv K(\omega,\sigma;\tau-\tau')$.
Functional differentiation gives the thermal average of a time-ordered
of position operators, i.e., the correlation functions for a stochastic 
process. For sake of 
completeness, we would like 
to present the simple result for the partition 
function of the anharmonic oscillator in the regime
$\lambda<<\omega$, in first order in $\lambda$. 
One finds \cite{livron}
\begin{equation}
Z(\beta)=\left(2\sinh(\frac{\beta\omega}{2})\right)^{-1}
\left(1-\frac{3\lambda}{4!}\int_{0}^{\beta}\,d\tau\frac{1}{4}
(\coth^{2}(\frac{\beta\omega}{2}))+O(\lambda^{2})\right). 
\label{11}
\end{equation}
To find the partition function for the anharmonic 
oscillator in the strong-coupling regime 
it is natural to use 
an unorthodox perturbative theory, i.e.,
the strong-coupling perturbative expansion.
The idea is to treat the Gaussian part of the action 
as a perturbation with respect to the 
non-Gaussian terms in the functional integral. 
We get the following formal representation 
for the generating functional at finite 
temperature $Z(\beta;h)$:
\begin{equation}
Z(\beta;h)=\exp\left(-\frac{1}{2}\int_{0}^{\beta} d\tau
\int_{0}^{\beta}d\tau'\frac{\delta}{\delta 
h(\tau)}K(\omega,\sigma;\tau-\tau')
\frac{\delta}{\delta h(\tau')}\right)\,Q(\beta,\sigma;\,h],
\label{12}
\end{equation}
where $Q(\beta,\sigma;h)$,
the new independent-value functional integral, is given by  
\begin{equation}
Q(\beta,\sigma;h)={\cal{N}}\int_{x(0)=x(\beta)}
[dx(\tau)]\,\exp\Biggl[
\int_{0}^{\beta} d\tau\,\left(-\frac{1}{2}\sigma\, 
\omega^{2}\,x^{2}(\tau)-
\,\frac{\lambda}{4!}x^{4}(\tau)
+h(\tau)x(\tau)\right)\Biggl],
\label{13}
\end{equation}
and the modified kernel $K(\omega,\sigma;\tau-\tau')$ was 
defined by Eq.(\ref{9}). The factor ${\cal{N}}$ is a 
normalization that can be found using that 
$Q(\beta,\sigma;h)|_{h=0}=1$.

The main difference from the standard representation for the 
generating functional is that 
we have an expansion of the generating functional
in inverse powers of the coupling constant. We are developing our 
perturbative expansion around the independent-value
generating functional $Q(\beta,\sigma;h)$.
We would like to stress that we are
considering a modification of the strong-coupling expansion.  
We split the quadratic part in the functional integral, 
which is proportional to the frequency squared, into two parts; 
one contributes together with the derivative term in 
the action as the perturbation, and 
the other appears in the independent-value generating functional.

One way to proceed is to neglect high-orders
terms in the perturbative expansion. Therefore in the 
leading order, we have that $Z(\beta;h)$ can be written as
\begin{equation}
Z(\beta;h)=\left(1-\frac{1}{2}\int_{0}^{\beta} d\tau
\int_{0}^{\beta}d\tau'\frac{\delta}{\delta 
h(\tau)}K(\omega,\sigma;\tau-\tau')
\frac{\delta}{\delta h(\tau')}\right)\,Q(\beta,\sigma;\,h),
\label{14}
\end{equation}
Since we are mainly interested in presenting the 
partition function, we can also assume that the external source is 
constant i.e. $h(\tau)=h$. As we will see, this assumption will lead us to redefine Klauder's representation for the independent-value 
generating functional. 
Of course, the above assumption considerably simplifies our problem, but
we still have some work to do.

To evaluate $\ln Z(\beta;h)$, note that we  have two steps to follow. The first one is 
to give meaning to the independent-value
generating-functional, and the 
second one is to regularize and renormalize the 
kernel $K(\omega,\sigma;\,\tau-\tau')$ integrated over the volume 
$[0,\beta]$. Note that the parameter $\sigma$ was introduced only
to simplify our 
calculations in some situations. Therefore $\sigma$ can be complex 
if we are able to work in all order of perturbation theory. 
The generating functional does not depends on the value for $\sigma$.
Since
we concentrate in the leading order, some care has to be taken to prevent 
a complex generating functional. 
A simple way to avoid the problem is assume that the parameter $\sigma$ is real. Therefore we will impose that  $\mbox{Im}\,\,(\sigma)=0$. 
In this situation
the independent-value functional $Q(\beta,\sigma;h)$ 
should be a normalized, positive definite functional. We will discuss
this point latter.

In the next section, we will use a combination of Klauder's
independent-value generating functional, and also use 
the generalized 
zeta-function method to regularize and renormalize the kernel $K(\omega,\sigma;\tau-\tau')$ integrated over the Euclidean 
time.

\section{The partition function for the 
anharmonic oscillator in the strong-coupling regime}\

In the present section we study the single anharmonic 
oscillator in the strong-coupling  regime. Since all the thermodynamics 
quantities are derived from the Helmholtz free energy, let us 
proceed in deriving the free energy.
The Helmholtz free energy 
can be obtained from $\ln Z(\beta,h)|_{h=0}$, i.e. $F(\beta)=-\frac{1}{\beta}\ln Z(\beta;h)|_{h=0}$.
To have a well defined
meaning to the Helmholtz free energy that can be 
obtained from Eq.(\ref{14}) we 
may proceed as follows.

First, we use the Klauder's result, as the 
formal definition of the independent-value
generating functional $Q(\beta,\sigma;h]$. Second, 
we have to regularize and renormalize the kernel 
$K(\omega,\sigma;\tau-\tau')$ integrated over the Euclidean 
time. The sucess of our method, depends critically on the possibility
to handle the independent-value generating functional and 
the kernel integrated on the Euclidean time $[0,\beta]$.

Since we are concerned with the strong-coupling regime, 
to evaluate $\ln Z(\beta)$ let us use the leading term.
Using the cumulant expansion idea, which relates the mean of a 
exponential to the exponential of means,  
after some simple calculations we obtain 
\begin{equation}
\ln Z(\beta;
h)=\frac{1}{Q(\beta,\sigma;h)}\frac{\partial^{2}Q(\beta,\sigma;
h)}{\partial\,h^{2}}
\left(-\frac{1}{2}+\frac{1}{2}
\frac{d}{ds}\zeta(s)|_{s=0}\right),
\label{27}
\end{equation}
where $\zeta(s)$ is the global generalized zeta-function associated with the operator $\left(-\frac{d^{2}}{d\tau^{2}}+
(1-\sigma)\,\omega^{2}\right)$. There are 
some issues that we would like to discuss.   
Note also that, we assume thermal equilibrium and
since we are working in the Euclidean formalism,
the spectrum of the operator $D\equiv\left(-\frac{d^{2}}{d\tau^{2}}+
(1-\sigma)\omega^{2}\right)$ has a denumerable contribution.
Remind that $\sigma$ is a complex parameter 
defined in the region $0\leq\,\mbox{Re}\,\,(\sigma)< 1$.
At this point, let us impose that $\mbox{Im}\,\,(\sigma)=0$, since 
$Q(\beta,\sigma;h)$ should be a positive definite functional, i.e., 
a characterized functional of a generalized stochastic process.

The operator $D$ is a positive definite eliptic
operator, and has a complete set of orthonormal eigenfunctions 
$x_{n}(\tau)$ and associated eigenvalues $a_{n}$. We have
\begin{equation}
\left(-\frac{d^{2}}{d\tau^{2}}+
(1-\sigma)\,\omega^{2}\right)x_{n}(\tau)=a_{n}\,x_{n}(\tau),
\label{28}
\end{equation}
with the boundary conditions $x_{n}(0)=x_{n}(\beta)$.
Note that we have
$\int_{0}^{\beta}d\tau\,x_{n}(\tau)x_{n'}(\tau)=\delta_{nn'}$.

At this point, we would like to stress that a number
of difficulties appear for complex $\sigma$.
Since we are working in the leading order of 
the inverse power of the coupling constant,
another problem related to the choice 
$\mbox{Im}\,\,(\sigma)\neq 0$, is related to the eigenvalue equation
given by Eq.(\ref{28}). For complex $\sigma$, the eigenvalue 
equation involve complex eigenvalues $a_{n}$ and the eigenfunctions
still form a complete, but not orthogonal set. See for example 
\cite{i1} \cite {i2}. Actually there is a program to study non-self
adjoint operators and this is related to the Parisi-Wu stochastic
quantization \cite{parisi} with a complex Langevin equation \cite{parisi2}
\cite{ambjornn}.
Actually complex Euclidean action also appears in different systems. 
For example the Euclidean action for a gauge theory with 
external static charge is complex. Effective actions with topological terms
are complex in the Euclidean formalism. 
Such kind of problems also appear in lattice QCD, where the analysis
of the phase diagram in the temperature-chemical potential plane has
been investigated \cite{ambjornn}. It is well known that in the 
SU(3) theory, the fermionic determinant is complex, giving to
a complex Euclidean action. For the study of the Gross-Neveu model 
with a nonzero imaginary chemical potential, see Ref. \cite{mar}.
Summarizing, complex Euclidean generating functional introduce new
stimulated problems and deserves a carefull analysis of them.

Going back to the eigenvalue equation,
the generalized zeta-function associated with 
the operator $\left(-\frac{d^{2}}{d\tau^{2}}+
(1-\sigma)\omega^{2}\right)$, i.e., $\zeta_{-\frac{d^{2}}{d\tau^{2}}+
(1-\sigma)\,\omega^{2}}(s)$ is defined by 
\begin{equation}
\zeta_{-\frac{d^{2}}{d\tau^{2}}+
(1-\sigma)\,\omega^{2}}(s)=\sum_{n=-\infty}^{\infty}a_{n}^{-s},
\label{29}
\end{equation}
where the spectrum is given by
\begin{equation}
a_{n}=\Biggl[\left(\frac{2\pi n}
{\beta}\right)^{2}+(1-\sigma)\,
\omega^{2}\Biggl],\,\,\,\, n\,\varepsilon\,\, \mbox{Z}.
\label{30}
\end{equation}
Using the definition for the global generalized 
zeta-function and the 
spectrum of the operator given by Eq.(\ref{30})
we have that the generalized zeta-function is given by
\begin{equation}
\zeta_{-\frac{d^{2}}{d\tau^{2}}+
(1-\sigma)\,\omega^{2}}(s)=\sum_{n=-\infty}^{\infty}
\Biggl[\left(\frac{2\pi n}
{\beta}\right)^{2}+(1-\sigma)\,\omega^{2}\Biggl]^{-s}.
\label{31}
\end{equation}
Here, it is useful to define the modified Epstein zeta-function in 
the complex plane $s$, i.e., the function $\zeta(s,\nu)$ by:
\begin{equation}
\zeta(s,\nu)=\sum_{n=-\infty}^{\infty}(n^{2}+\nu^{2})^{-s},
\,\,\,\,\,\,\,  \nu^{2}\,\,>0. 
\label{32}
\end{equation}
The series defined by Eq.(\ref{32}) 
converges absolutely and defines in the complex $s$ plane an 
analytic function
for $\mbox{Re(s)}>\frac{1}{2}$. It is possible to analytically
extend the modified Epstein zeta-function where the integral
representation is valid for $Re(s)<1$, \cite{ford} \cite{fn}:
\begin{equation}
\sum_{n =-\infty}^{\infty} \bigl( n^{2} +\nu^{2} \bigr)^{-s}
= \nu^{1-2s} \Biggl[ \sqrt{\pi}\, {{\Gamma(s-{1\over 2})} \over
{\Gamma(s)}} 
+ 4 \sin{\pi s} \int_1^{\infty} 
{{(t^{2}-1)^{-s} dt} \over {{\rm e}^{2\pi\nu t} -1}} \Biggr] \, . 
\label{33}
\end{equation}
For a different representation for the analytic
extention of the modified Epstein zeta-function in terms of the
modified Bessel function $K_{\alpha}(z)$ or the Macdonald's function, the reader can use for example Ref. \cite{B}.

It is not difficult to write the generalized zeta-function in 
terms of the modified Epstein zeta-function. We have 
\begin{equation}
\zeta_{-\frac{d^{2}}{d\tau^{2}}+
(1-\sigma)\,\omega^{2}}
(s)=\left(\frac{\beta}{2\pi}\right)^{2s}\zeta\left(s,
\sqrt{1-\sigma}(\frac{\omega\beta}{2\pi})\right),
\label{34}
\end{equation}
where $\zeta(s,\nu)$, is the modified Epstein
zeta-function. As we discussed, the series representation 
for $\zeta(s,\nu)$ converges
for $\mbox{Re(z)}>\frac{1}{2}$ and its analytic continuation defines a meromorphic
function of $s$ which is 
analytic at $s=0$. The modified Epstein
zeta-function has poles at $s=\frac{1}{2},-\frac{1}{2}$, etc.
Using Eq.(\ref{33}) is not difficult to show that  the values for the modified Epstein zeta-function $\zeta(s,\nu)$, at $s=0$ and $\frac{\partial}{\partial s}\zeta(s,\nu)|_{s=0}$ are given by 
\begin{equation}
\zeta(s,\nu)|_{s=0}=0,
\label{35}
\end{equation}
and also
\begin{equation}
\frac{\partial}{\partial s}\zeta(s,\nu)|_{s=0}=-2\ln(2\,\sinh\pi\nu).   
\label{36}
\end{equation}
Since we are interested in calculating
the derivative of the generalized zeta-function at the origin 
of the complex $s$ plane, we have
\begin{equation}
\frac{1}{2}\frac{\partial}{\partial s}\zeta_{-\frac{d^{2}}{d\tau^{2}}+
(1-\sigma)\,\omega^{2}}(s)|_{s=0}=
\left(\frac{1}{2}\zeta(s,\nu)
\frac{d}{ds}\left(\frac{\beta}{2\pi}\right)^{2s}
+\frac{1}{2}\left(\frac{\beta}{2\pi}\right)^{2s}
\frac{\partial}{\partial s}\zeta(s,\nu)\right)|_{_{s=0}}.
\label{37}
\end{equation}
Choosing $\sigma=0$, and 
using Eq.(\ref{35}) and Eq.(\ref{36}) in Eq.(\ref{37}) we obtain the  
well known result in the literature. For the general case $(\sigma\neq 0)$
we have
\begin{equation}
\frac{1}{2}\frac{\partial}{\partial s}\zeta
_{-\frac{d^{2}}{d\tau^{2}}+
(1-\sigma)\,\omega^{2}}(s)|_{s=0}=
-\ln\Biggl[(2\,\sinh\left((1-\sigma)\frac
{\omega\beta}{2}\right)\Biggl].
\label{ii}
\end{equation}
Thus it is clear that the zeta function regularization  
can be used to control the divergences of the kernel 
$K(\omega,\sigma;\,\tau-\tau')$ integrated over the Euclidean time.

We would like to stress that 
we are using the Klauder's result, as the 
formal definition of the independent-value generating functional
derived for scalar fields in a $d$-dimensional Euclidean space.
We would like to point out that
in Klauder's derivation for the independent-value
model a result was obtained which is well defined for all 
functions which are square integrable in $R^{n}$ i.e.,
$h(x)\,\, \varepsilon\,L^{2}(R^{n})$. 
This observation allow us to conclude that we need also to
use a normalization in the situation that we are investigating.
It is possible to show that the independent-value 
generating function can be written as
\begin{equation}
Q(\beta,\sigma;h)=\exp\Biggl[-\frac{1}{2\beta}\int_{0}^{\beta}d\tau
\int_{-\infty}^{\infty}\frac{du}{|u|}\left(1-\cos(hu)\right)\exp
\left(-\frac{1}{2}\,\sigma\,\omega^{2}\,u^{2}-
\frac{\lambda}{4!}\,u^{4}\right)\Biggl].
\label{353}
\end{equation}
There is no need to go into details of this derivation. 
The reader can find it
in Ref. \cite{jrk} \cite{sol1} \cite{kla2}. 
It is important to stress that the  
independent-value generating function defined 
by Eq.(\ref{353}) do not reduce to the conventional free 
independent-value model, if we choose 
$\lambda=0$. The non-gaussian contribution represents a 
discontinuous perturbation of the free theory, or using the 
Klauder's definition; a pseudo-free theory. 
Actually, this is the key point of the program developed by Klauder to investigate non-renormalizable models in field theory. For 
a interesting study of the pseudo-free harmonic oscillator see for
example Ref. \cite{pseudo}.

In order to study $Q(\beta,\sigma;h)$, let us 
define $E(\omega,\sigma,\lambda;h)$ given by
\begin{equation}
E(\omega,\sigma,\lambda;h)= 
\int_{-\infty}^{\infty}\frac{du}{|u|}\left(1-\cos(hu)\right)\exp\left(
-\frac{1}{2}\,\sigma\,\omega^{2}\,u^{2}-
\frac{\lambda}{4!}\,u^{4}\right).
\label{aa}
\end{equation}
Using a series representation for $\cos x$,
it is not difficult to show that
\begin{equation}
E(\omega,\sigma,\lambda;h)=
2\sum_{k=1}^{\infty}\frac{(-1)^{k}}{(2k)!}h^{2k}
\int_{0}^{\infty}du\,u^{2k-1}\exp
\left(-\frac{1}{2}\,\sigma\,\omega^{2}\,u^{2}-
\frac{\lambda}{4!}\,u^{4}\right).
\label{38}
\end{equation}
Now let use the fact that the $\sigma$ parameter can be choosen in such 
a way that the calculations becomes tractable. Analysing
only the independent-value generating functional it is not possible 
to write $Q(\beta,\sigma;h)$ in a closed form even in the case of 
constant external source. 
One way to obtain a closed expression is to choose $\sigma=0$. 
Therefore we have
\begin{equation}
E(\omega,\sigma,\lambda;h)|_{\sigma=0}=
2\sum_{k=1}^{\infty}\frac{(-1)^{k}}{(2k)!}h^{2k}
\int_{0}^{\infty}du\,u^{2k-1}\exp(-\frac{\lambda}{4!}\,u^{4}).
\label{39}
\end{equation}
At this point let us use the following 
integral representation 
for the Gamma function \cite{grads}
\begin{equation}
\int_{0}^{\infty}\,dx\,x^{\nu-1}\exp(-\mu x^{p})=\frac{1}{p}
\mu^{-\frac{\nu}{p}}\,\Gamma\left(\frac{\nu}{p}\right),\,\,\,\,\,
\mbox{Re}(\mu)>0\,\,\,\,\,\,\,\mbox{Re}(\nu)>0\,\,\,\,p>0.
\label{40}
\end{equation}
It is clear that the 
$(\lambda\,x^{p})$ theory, for even $p>4$, can also easily 
handle applying our method. Using the result given by Eq.(\ref{40}) in Eq.(\ref{39})  we have
\begin{equation}
E(\omega,\sigma,\lambda;h)|_{\sigma=0}=\sum_{k=1}^{\infty}
g(k)\frac{h^{2k}}{\lambda^\frac{k}{2}},
\label{41}
\end{equation}
where the coefficients $g(k)$ are given by
\begin{equation}
g(k)=\frac{1}{2}\frac{(-1)^{k}}{(2k)!}(4!)^
{\frac{k}{2}}\Gamma(\frac{k}{2}).
\label{42}
\end{equation}
Substituting the Eq.(\ref{41}) and Eq.(\ref{42}) in Eq.(\ref{353}) we obtain
that the independent-value generating function
$Q(\beta,\sigma;h)|_{\sigma=0}$ can be written as 
\begin{equation}
Q(\beta,\sigma;h)|_{\sigma=0}=\exp\Biggl
[-\frac{1}{2\beta}\int_{0}^{\beta}
d\tau\,\sum_{k=1}^{\infty}\,g(k)\frac{h^{2k}}{\lambda^\frac{k}{2}}\Biggl].
\label{43}
\end{equation}
It is easy to calculate the second derivative for 
the independent-value generating function with respect to $h$. Note that 
$Q(\beta,\sigma;h)|_{_{h=\sigma=0}}=1$. Thus we have  
\begin{equation}
\frac{\partial^{2}Q(\beta,\sigma;
h)}{\partial\,h^{2}}|_{\sigma=0}=
\left(-\frac{1}{2}
\sum_{k=1}^{\infty}\,g(k)(2k)(2k-1)\frac{h^{2k-2}}
{\lambda^\frac{k}{2}}\right)
\exp\left(-\frac{1}{2}
\sum_{k=1}^{\infty}\,g(k)\frac{h^{2k}}{\lambda^\frac{k}{2}}\right)
+G(h),
\label{44}
\end{equation}
where $G(h)$ is given by
\begin{equation}
G(h)=\left(\sum_{k,\,q
=1}^{\infty}\,g(k,q)\frac{h^{2k+2q-2}}{\lambda^\frac{k+q}{2}}\right)
\exp\left(-\frac{1}{2}
\sum_{k=1}^{\infty}\,g(k)\frac{h^{2k}}{\lambda^\frac{k}{2}}\right),
\label{iii}
\end{equation}
and $g(k,q)=k\,q\,g(k)g(q)$. We are interested in the case $h=0$, 
therefore 
the double series does not contributes to the Eq.(\ref{44}), since $lim_{h\rightarrow 0} G(h)=0$.
Using the fact that we
are interested in the case $h=0$, we have the simple result that in the 
Eq.(\ref{44}) only the term $k=1$ contributes.
We get 
\begin{equation}
\frac{\partial^{2}Q(\beta,\sigma;
h)}{\partial\,h^{2}}|_{h=\sigma=0}=\sqrt{\frac{3\pi}{8\lambda}}.
\label{45}
\end{equation}
Substituting the result obtained from the generalized 
zeta-function method
given by Eq.(\ref{ii}) (choosing $\sigma=0$) 
and Eq.(\ref{45}) in Eq.(\ref{27}) we have that 
$\ln Z(\beta)$ is given by 
\begin{equation}
\ln Z(\beta)=\sqrt{\frac{3\pi}{8\lambda}}\Biggl[\frac{1}{2}-
\ln\left(2\sinh(\frac{\omega\beta}{2})\right)\Biggl].
\label{451}
\end{equation}
Therefore the partition function for the single oscillator is 
\begin{equation}
Z(\beta)=\frac{e^{\frac{1}{2}\sqrt{\frac{3\pi}{8\lambda}}}
}{\left(2\sinh(\frac{\omega\beta}{2})\right)^{
\sqrt{\frac{3\pi}{8\lambda}}}}.
\label{46}
\end{equation}
It should be noted that for a system of $N$ harmonic oscillators, the 
partition function is 
\begin{equation}
Z(\beta)=\frac{1}{\left(2\sinh(\frac{\omega\beta}{2})\right)^{N}}.
\label{nova}
\end{equation}
Therefore the identification $N=
\sqrt{\frac{3\pi}{8\lambda}}$, the partition function for the 
strongly coupled single oscillator seem to be proportional to the 
partition function of $N$ harmonic oscillators, up to the order $\frac{1}{\sqrt{\lambda}}$. 

Other thermodynamics
quantities that we are able to find are the Helmholtz
free energy and the mean energy. 
The Helmholtz free energy is given by $F(\beta)=-\frac{1}{\beta}\ln Z(\beta,h)|_{h=0}$. Thus we have
\begin{equation}
F(\beta)=\frac{1}{\beta}\sqrt{\frac{3\pi}{8\lambda}}
\Biggl[-\frac{1}{2}+
\ln\left(2\sinh(\frac{\beta\omega}{2})\right)\Biggl].
\label{461}
\end{equation}
It is possible to write this expression in the following way
\begin{equation}
F(\beta)=\sqrt{\frac{3\pi}{8\lambda}}\Biggl[
-\frac{1}{2\beta}+\frac{\omega}{2}+
\frac{1}{\beta}\ln\left(1-e^{-\beta\omega}\right)\Biggl].
\label{fe}
\end{equation}
Finally the mean energy is defined by $E=-\frac{\partial}{\partial\beta}
\ln\,Z(\beta)|_{h=0}$. Therefore we have
\begin{equation}
E=\sqrt{\frac{3\pi}{8\lambda}}
\Biggl[\frac{\omega}{2}+\frac{\omega}{e^{\omega\beta}-1}\Biggl].
\label{462}
\end{equation}
We conclude with some observations.
Usually, the free energy of a finite system is an analytic 
function of the parameters that define our physical system. 
Nevertheless, our results are showing that the thermodynamics
quantities are nonanalytic in the coupling constant.

The picture emerging from the previous 
discussion is the following: in the 
strong-coupling perturbative expansion we may 
split the problem of defining the 
generating functional into two parts: how to define precisely
the independent-value generating functional and how 
to go beyond the independent-value approximation, taking into account the perturbation part. 
Our results show that the strong-coupling perturbative 
expansion, in combination with an analytic regularization 
procedure, is a useful method to compute global quantities, as
the Helmholtz free energy, in the strong-coupling regime.

\section{Conclusions}

In this article we studied the strong-coupling regime 
in one-dimensional models, after analytic continuation to imaginary time.
One-dimensional models are very simple system for which we can 
apply our method in obtaining thermodynamics quantities in
the leading order in the inverse of coupling constant.
We calculate the partition function and the Helmholtz free energy 
for the anharmonic oscillator, using the strong-coupling perturbative 
expansion and the generalized zeta-function analytic regularization.
It was possible to present expressions up to the 
order $\frac{1}{\sqrt{\lambda}}$ for 
the partition function and the other thermodynamic quantities derived 
from the the Helmholtz free energy.

The
picture that emerges from our method is the following: in the 
strong-coupling perturbative expansion we may 
split the problem of defining the 
generating functional into two parts: the first is how to define precisely
the independent-value generating functional. We may use of a lattice approximation to give a mathematical 
meaning to the non-Gaussian functional. Actually, it is not easy 
to recover the continuum limit, and as we discussed the use of the 
central limit theorem leads us to a Gaussian theory. 
Instead of this, we are using the Klauder's
independent-value generating functional.
This is the key point of the article and can explain the discrepance
between our results and the Ref. \cite{livron} (pp.930,931), where 
the large coupling behavior of the anharmonic oscillator is discussed.

The second part is to go beyond the independent-value  
approximation and take into account the perturbation part. 
This problem can be controled using 
an analytic regularization. Besides these technical problems, 
we still have the problem of obtaining the Green's
functions from this approach. The strong-coupling perturbative 
expansion is not fit to obtain local quantities, as  
the Green's functions of the model. 
On the other hand, our results show that the strong-coupling perturbative 
expansion, in combination with an analytic regularization 
procedure, is a useful method to compute global quantities, as
the Helmholtz free energy, in the strong-coupling regime.

There are several directions for investigations, using the 
Klauder's result and an analytic regularization procedure. 
It should be possible for applying the method to more 
realistic theories. To mention a few: since scalar fields play a 
fundamental role in the standard model, the study of 
the strongly coupled $(\lambda\varphi^{4})_{d}$
theory at finite temperature and also the renormalized vacuum energy
of scalar quantum fields in the presence of macroscopic structures 
deserves future investigations.

\section{Acknowlegements}

I would like to thanks, G. F. Hidalgo and S. Joffily for enlightening
discussions, ans also L. A. Oliveira for comments on the manuscript.
This paper was supported by Conselho Nacional de
Desenvolvimento Cientifico e Tecnol{\'o}gico do Brazil (CNPq).

\end{document}